\documentclass[conference]{IEEEtran}
\IEEEoverridecommandlockouts
\usepackage{cite}
\usepackage{amsmath,amssymb,amsfonts}
\usepackage{algorithmic}
\usepackage{graphicx}
\usepackage{textcomp}
\usepackage{amsmath}
\usepackage{amssymb}
\usepackage{gensymb}
\usepackage{xcolor}
\usepackage{color,soul}
\usepackage{makecell}
\usepackage{multirow}
\usepackage{colortbl}
\usepackage{hhline}
\usepackage{subcaption}
\def\BibTeX{{\rm B\kern-.05em{\sc i\kern-.025em b}\kern-.08em
    T\kern-.1667em\lower.7ex\hbox{E}\kern-.125emX}}
\begin{document}

\title{A Controlled Study on Evaluation of Thermal Stimulation Influence on Affective Measures of Uninformed Individuals\\
{}
\thanks{Authors $^1$ are with the Department of Mechanical Engineering at University of South Florida. Authors $^2$ are with the Department of Psychology at University of South Florida. }
}

\author{\IEEEauthorblockN{Mehdi Hojatmadani{$^1$}}

\and
\IEEEauthorblockN{Samantha Shepard{$^2$}}

\and
\IEEEauthorblockN{Kristen Salomon{$^2$}}

\and
\IEEEauthorblockN{Kyle Reed{$^1$}}

}

\maketitle

\begin{abstract}
Although the relationship between temperature and emotional states has been investigated in the field of haptics, it remains unknown if, or in what direction, temperature affects emotional states. We approach this question at the intersection of haptics and psychology using a custom-built thermal device and emotional responses based on photos from the International Affective Picture System (IAPS) library. Unlike past research, this study incorporates deception and a control (i.e., neutral temperature) condition. One hundred and twenty naive subjects reported their emotional responses to fifty-six images varying on normative arousal and valence ratings while being exposed to a cool~(30\degree C), neutral (33\degree C), or warm (36\degree C) temperature applied to the upper back. Participants exposed to warm temperatures reported higher arousal ratings in some image categories than participants exposed to neutral or cool temperatures. Valence ratings were decreased in warm conditions compared to neutral conditions. The emotion wheel was used as a complementary method of affective response measurement, and exploratory analysis methods were implemented. Although the valence and arousal showed statistical significance, the emotion wheel results did not demonstrate any significant differences between the temperature conditions.
\end{abstract}


\section{Introduction}
Temperature plays a vital role in human survival. Thermoregulation attunes body temperature to the environment, utilizing mechanisms such as sweating or shivering~\cite{liedtke2017deconstructing}. Temperature has long been posited to play a role in emotions~\cite{ziegler1938study}. Most research has focused on the effect of emotional states on body temperature, while less work has examined the effects of body temperature on resultant emotional states. The present study examines the effects of local thermal stimulation on emotional experiences.

Evidence suggests that people can communicate emotions such as anger, fear, and love through physical touch~\cite{hertenstein2006touch}. Similarly, greater emotional pleasantness and excitement are associated with less roughness and more fluidity while interacting with objects through touch~\cite{drewing2017feeling}. Given these findings, it seems likely that affective measures may also be associated with thermal sensation~\cite{barbosa2021temperature}.

Emotions can be described quantitatively based on two measures: valence and arousal. Positive/negative valence describes the pleasantness/unpleasantness of an emotion, while high/low arousal describes the exciting/calming degree of the emotion. For example, anger is an emotion with negative valence and high arousal, while feeling peaceful is associated with positive valence and low arousal.

\subsection{Effect of emotional experience on body temperature}

The effect of emotions on body temperature is well established.  Stress is associated with small increases in central body temperature, referred to as stress-induced hyperthermia~\cite{oka2018stress}. Further, specific emotional states are associated with changes in skin temperature~\cite{ekman1983autonomic,levenson1990voluntary}, which is thought to reflect changes in cutaneous circulation~\cite{elam1987skin}. Given these bodily changes, it is perhaps not surprising that individuals associate specific emotions with distinct bodily patterns of warm and cool, or ``temperature maps"~\cite{nummenmaa2014bodily}. For example, anger is associated with heat in the head, torso, and arms, whereas anxiety is associated with heat mainly in the chest and warmth in the neck and gut.  Sadness is also associated with warmth in the face, neck, and chest but coolness in the arms and legs. In contrast, depression is associated only with coldness in the head and extremities.  Thus, emotional experience involves changes in body temperature and sensations of temperature, which suggests that changes in body temperature may affect emotional experience. 

\subsection{Effect of Temperature Changes on Emotional Experience}

Temperature's effects have been examined as either the emotional interpretation of the temperature itself or as the effects of temperature on broader emotional experience. Regarding the former, warm stimuli are generally perceived as positive in emotional valence, and cool stimuli are perceived as unpleasant in valence~\cite{wilson2017multi}. Evidence is mixed regarding the latter and the focus of the present study.

A study of the effect of a combination of haptic and visual stimuli on altering emotional ratings demonstrated that temperature can increase the available range of emotional states~\cite{halvey2012augmenting}. Warm and cool temperatures effectively communicate positive and negative meanings and improve the quality of communicated emotions~\cite{suhonen2012haptically}. It was demonstrated that warmth stimulation increases emotional arousal while cool stimulation affects perceptions of the pleasantness of the message, also known as valence. For example,~El Ali et al.~\cite{el2020thermalwear} played a series of emotionally neutral audio messages to participants while wearing a thermally active vest, applying a temperature gradient on their bodies. The temperature changes in positive (warm) and negative (cool) directions influenced the valence and arousal levels of the messages they heard. Warming temperature has been shown to elicit greater emotional arousal than cooling temperatures, while individuals viewed images that elicit specific emotions~\cite{akazue2017using,akazue2016effect}. It was also demonstrated that a 3\degree C temperature change increased arousal levels compared to no temperature change.

Temperature perception can be manipulated to create various sensations and emotions, for instance, creating a constant cooling sensation~\cite{hojatmadani2018asymmetric,manasrah2020computational,hojatmadani2021interrelation} when the average temperature of the stimulated area is not changing or creating a pain sensation~\cite{craig1994thermal} using a combination of hot and cold bars where the temperature of each bar is below the pain threshold. Temperature is also useful in modulating affective states  when combined with other sensory modalities. Additionally, therapeutic manipulation of brain temperature is hypothesized to decrease depressed mood and may be effective in treating psychological disorders, such as mania~\cite{salerian2008brain}. A recent study of whole-body cooling and warming found resultant activation not only in the thermosensory cortex (left dorsal posterior insula) but also in limbic structures (the left amygdala and the bilateral retrosplenial cortices) associated with autonomic function and emotion~\cite{oi2017neural}.

One of the instances of the effect of temperature on emotions was demonstrated by a test designed by De Wall and Bushman~\cite{dewall2009hot} in which aggression was evaluated after being exposed to words related to temperature. The results demonstrated that the words associated with heat could invoke aggressive thoughts and hostile perceptions, which reflect high-arousal emotion. Janssen et al.~\cite{janssen2016whole} studied the effect of temperature on depression to determine whether thermal stimulation could be used as a treatment method for major depressive disorder. The results indicated that a 6-week hyperthermia intervention of increasing the body temperature decreased the depression symptoms relative to a control condition. Another study~\cite{hanusch2013whole} demonstrated that whole-body hyperthermia treatment reduced depression in the participants who were not given SSRI medications.  Notably, deception was not used in these experiments, which could have influenced participant responses. Participants were aware that a thermal stimulus was present and likely could have deduced that the experiment was testing if temperature could influence their emotions. 

Although the previously discussed research reveals a strong connection between temperature and emotion, the efficacy of local thermal stimulation needs to be thoroughly investigated.

\section{The Present Study}

Our study extends extant research by investigating how temperatures (warm or cool) may affect the intensity and pleasantness of emotion by applying thermal stimulus to local areas of the body. In this study, the thermal actuator was in contact with the skin on the back of the neck. Since the neck does not typically experience external thermal stimuli, unlike hands, and temperature sensation is not as localizable as other senses, it is hypothesized that the sensation of the temperature will be more pronounced in this area. Another distinction of this work from similar studies is that participants were not aware of the study's purpose in this study, making the results less susceptible to expectation bias. We hypothesize that placing the thermoelectric device on the back of the neck reduced the possibility of participants realizing the purpose of the study since they were not required to move their hands on/off the device actively. Another improvement in this study is the addition of a control condition, where participants' ratings of images in cool and warm conditions were compared to the neutral temperature condition. This is in contrast to the existing studies where participants' valence and arousal ratings of each image were compared to the rated IAPS library ratings~\cite{akazue2016effect}. 

We hypothesize that the temperature condition will affect the SAM valence and arousal rating for the images in warm and cool conditions compared to the neutral condition. Participants will also be asked to report their emotions using Geneva Emotion Wheel as a complementary method of reporting their affective state.

\section{Methods}

\subsection{Participants}

Participants were recruited from the undergraduate students participant pool in the Department of Psychology at the University of South Florida.  
One hundred and twenty-nine individuals participated in the study. The experiments were completed between August 2019 and November 2019. Nine participants' data were dismissed because the correct temperature was not applied due to equipment malfunction, leaving 41 participants in the cool condition, 37 in the warm condition, and 42 in the neutral condition.

The experimental protocol and the consent form were approved by the University of South Florida's 
Institutional Review Board. Written informed consent was obtained from all the subjects prior to participation. All research was performed in accordance with the relevant guidelines and regulations.

\subsection{Experimental Setup}

\begin{figure}
	\centering
	\includegraphics[width=3.4in]{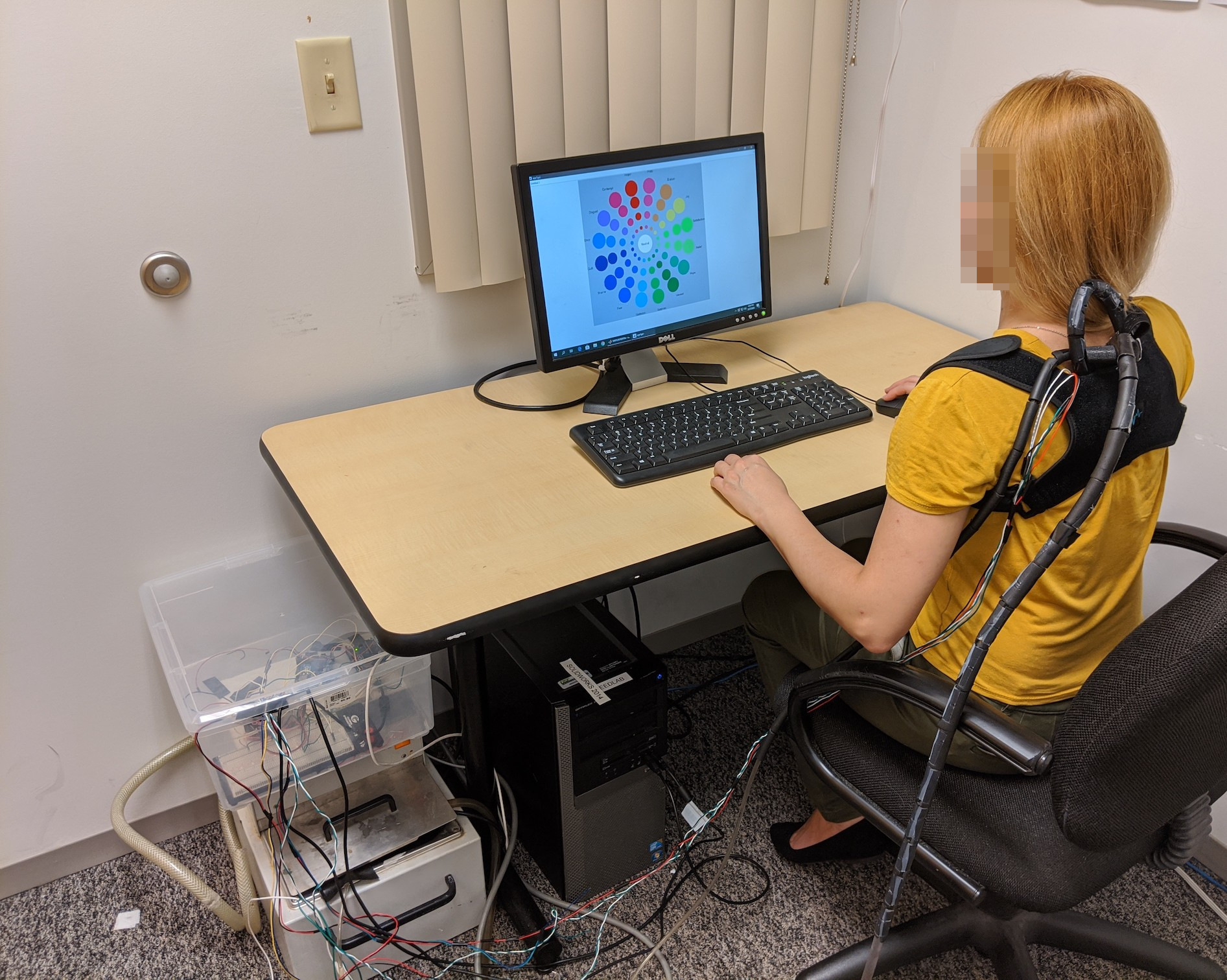}
	\caption{Experimental setup. A participant wearing the thermal actuator on the back of the neck while rating the image using the Geneva Emotion Wheel.}
	\label{fig:setup}
\end{figure}

The experimental setup (Figure~\ref{fig:setup}) is comprised of three main components: the thermoelectric device, the temperature controller, and the graphical user interface. In this study, a Tegway flexible thermoelectric device~\cite{kim2018structural} was utilized to create temperature gradients. In order to control the skin temperature during the experiment, the temperature of the other side of the thermoelectric needs to be maintained at a low temperature. An active cooling system was designed to be attached to the thermoelectric device for this purpose. A system of water block and hoses was used to supply cold water to the hot side of the device. The thermal device was held in place using a wearable brace, securing the device's position. The thermoelectric temperature at the skin interface was measured using an NTC 10K ohm thermistor. The power supply to the thermoelectric was controlled using a Pololu Simple Motor Controller 18v7 and an Arduino Uno microcontroller. The actuator was able to apply thermal cues with an accuracy of $\pm$0.5~{\degree}C. The motor controller was powered through a power supply (B$\&$K Precision 1666). A graphical user interface (GUI) was developed in MATLAB to display the images and record the affective responses. MATLAB was communicating with the Arduino through a serial port to apply the thermal cues and automate the experiment.

\subsection{Study Design}

Participants viewed a series of images and self-reported their affective response to each using the graphical user interface. Participants were randomly assigned to experience one of three temperature conditions. This approach was adapted to avoid raising participants' suspicion about the study's goal, as this was the case in a pilot study where temperature condition was used as a within-subject factor. Images were selected based on the valence and arousal ratings of the IAPS normative rates such that the standard deviation of the means for the selected images was minimized. Each participant viewed eight images in each of the seven image categories.

\subsection{Material}

\textit{Affective Stimuli:} Photos from the International Affective Picture System (IAPS; \cite{lang1997international}) were shown to participants. The IAPS is a pictorial library of images that have been demonstrated to elicit specific emotions. Each picture is accompanied by a normative mean valence and arousal level. We categorized pictures into nine categories based on three levels of low, medium, and high valence and arousal normative scores. These levels were selected to create the most distinct set of images from each other. However, given the selection criteria, two categories did not contain a sufficient number of images (negative valence-low arousal, N~=~5 images, and neutral valence-high arousal, N~=~3 images). Thus, a subset of eight images was selected from the remaining seven categories. To create relatively homogenous categories, images for each category were selected to minimize the standard deviation of ratings within each category using a custom MATLAB optimization script. The depiction of each image category location on the valence-arousal plane and the normatively rated valence and arousal levels are shown in Table~\ref{tbl:normative responses}. 

\begin{table}[b]
\centering
\caption{Normative rating of valence and arousal in each image category (top) and mean of participants' response for those categories (bottom).}
\resizebox{\columnwidth}{!}{
\begin{tabular}{|c|c|c|c|c|} 
\hline
\multicolumn{1}{|c}{} &                                                                      & \multicolumn{1}{c}{-}                                                  & \multicolumn{1}{c}{Valence}                                            & +                                                                              \\ 
\hline
\tiny{High}                     & \begin{tabular}[c]{@{}c@{}}Normative Rating\\Responses~\end{tabular} & \begin{tabular}[c]{@{}c@{}}{[}2.18, 6.32]\\{[}2.09, 5.53]\end{tabular} & {\cellcolor[rgb]{0.502,0.502,0.502}}                                   & \begin{tabular}[c]{@{}c@{}}{[}7.33, 6.00]\\{[}6.57, 4.63]\end{tabular}         \\ 
\cline{2-5}
Arousal               & \begin{tabular}[c]{@{}c@{}}Normative Rating\\Responses~\end{tabular} & \begin{tabular}[c]{@{}c@{}}{[}2.93, 5.03]\\{[}3.09, 4.42]\end{tabular} & \begin{tabular}[c]{@{}c@{}}{[}5.12, 4.49]\\{[}5.23, 3.61]\end{tabular} & \begin{tabular}[c]{@{}c@{}}{[}7.53, 4.60]\\{[}6.93, 3.80]\end{tabular}         \\ 
\hhline{|~----|}
\tiny{Low}                     & \begin{tabular}[c]{@{}c@{}}Normative Rating\\Responses~\end{tabular} & {\cellcolor[rgb]{0.502,0.502,0.502}}                                   & \begin{tabular}[c]{@{}c@{}}{[}5.21, 3.43]\\{[}5.18, 3.30]\end{tabular} & \begin{tabular}[c]{@{}c@{}}{[}7.3, 3.61]\\{[}6.75, 3.30]
\end{tabular}  \\
\hline
\end{tabular}
}

\label{tbl:normative responses}
\end{table}

\textit{Ratings of IAPS photos:} The self-assessment Manikin (SAM; \cite{bradley1994measuring}) was used to collect participants' valence and arousal ratings on a nine-point scale. SAM ranges from a smiling, happy figure to a frowning, unhappy figure when representing the valence dimension and ranges from an excited, wide-eyed figure to a relaxed, sleepy figure for the arousal dimension. The SAM also includes a dominance rating scale. However, because valence and arousal can explain most of the variations in emotions~\cite{russell1980circumplex}, the dominance rating was not used in the data collection process. 

\textit{Geneva Emotion Wheel:} Emotional responses were collected using the Geneva emotion wheel~\cite{sacharin2012geneva}. This emotion wheel is comprised of 16 emotions forming a circle with neutral in the middle. The concept of the emotion wheel is that the further away from the center of the circle, the more intense the emotions become. Therefore, participants select the biggest circle on the joy axis if they feel much joy. The emotion wheel could be considered a more direct report method as there are distinct emotions on the wheel versus SAM where emotions are broken into their components.

\subsection{Procedure}

Participants were informed that the study aimed to evaluate emotion while recording their physiological responses. They were briefed about the procedure of the study. This provided adequate time for their body to acclimate to the room  temperature. After the informed consent was signed and briefing was completed, the thermoelectric device was placed on the back of the neck approximately at the C7 to T2 vertebrae level. Research personnel placed two ECG electrodes onto the collarbone and torso, where the electrodes were attached to physiological recording equipment. However, no physiological measures were collected. The physiological recording equipment served as a distraction from the study's true purpose, as participants were led to believe the thermoelectric device was yet another physiological recording device.

Participants sat in a chair in front of a computer and were given instructions on how to complete the self-assessment manikin and the emotion wheel.  Participants were led through several practice trials to ensure they understood the emotion rating procedure. Next, each participant was randomly assigned to one of the three temperature conditions: cool (30\degree C), neutral (33\degree C), and warm (36\degree C). Participants viewed each of the 56 selected IAPS images for 6 seconds, displayed in random order on the computer screen, and were asked to rate their emotions using the SAM and the emotion wheel. The temperature was constant and was applied throughout the experiment. The trial practice and the main experiment lasted approximately 35 minutes.

Once all images were viewed and rated, participants were probed for suspicion of the cover story, fully debriefed as to the nature of the deception, and asked if, after knowing the true nature of the study, they would like to have their data removed or to allow the researcher to retain it for analysis (in accordance with the university's IRB policy). No subject requested their data to be removed. No subject realized the true nature of the study. 

\section{Results}

Two mixed-design factorial 3 (Temperature) x 7 (image category) ANOVA tests were conducted with temperature as the between-subjects variable, image category as the within-subjects variable, and image ratings as the dependent variable. Bonferroni-corrected post-hoc comparisons were completed when omnibus F-tests showed statistical significance. Statistical significance was determined using an alpha of 0.05. Analyses were conducted using SPSS 24. The single asterisk in the figures represents 0.01~$<$~p~$<$~0.05.

\subsection{Self-Assessment Manikin}

\begin{figure*}
\centering
\includegraphics[width=\textwidth ]{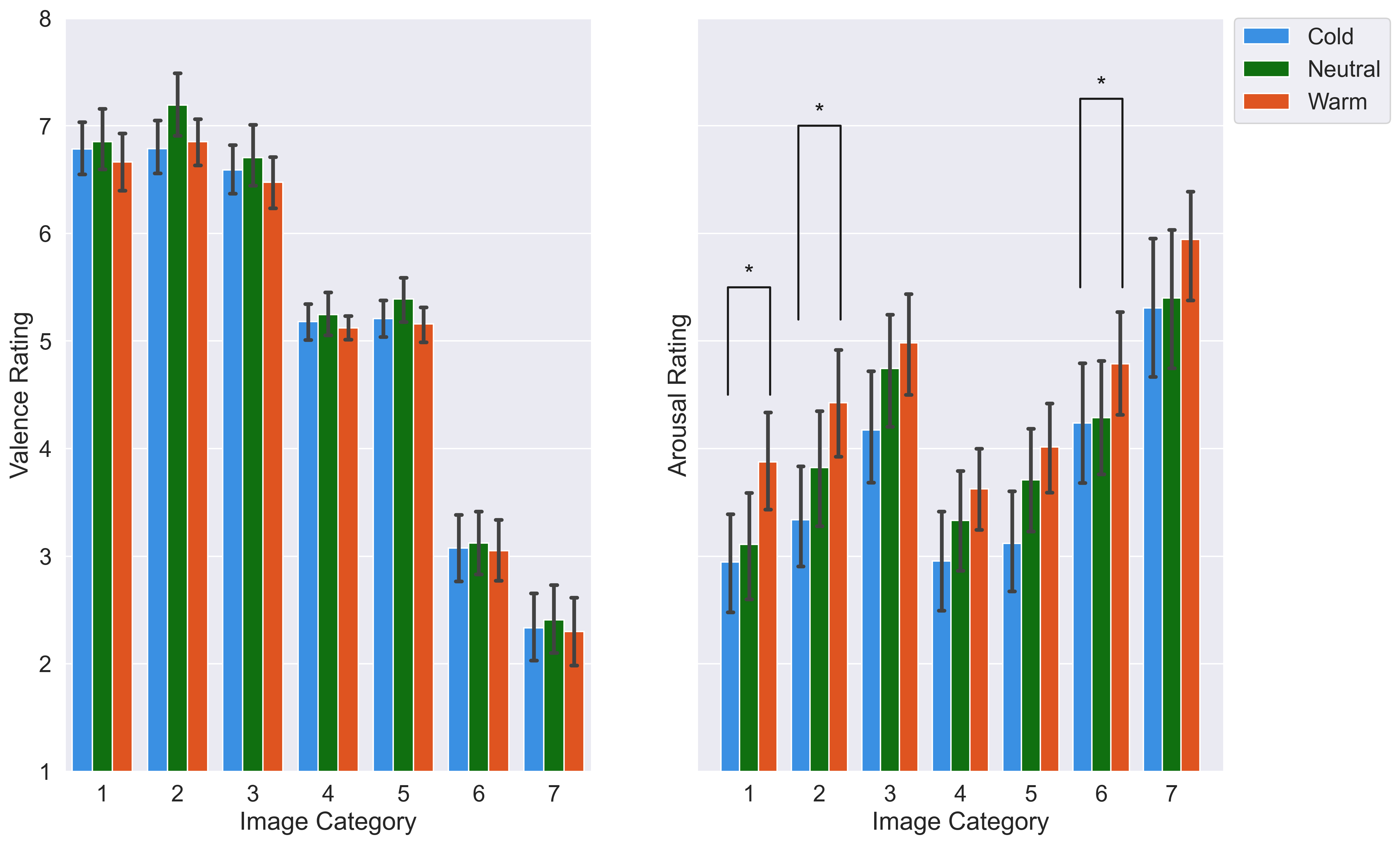}
\caption{SAM rating responses for valence and arousal. Statistical significance is demonstrated on the plots with the asterisks. The error bars represent the standard error values.}
\label{fig:valence_arousal}
\end{figure*}

The ANOVA test results for arousal ratings demonstrated a main effect of temperature (F(2,116) = 3.10, p = 0.048). Arousal ratings across the image category were significantly lower in the cool condition (M = 3.75, SE = 0.21) than in the warm condition (M = 4.52, SE = 0.22). These two conditions were not rated statistically different from the control (neutral) condition. Post-hoc tests indicated that participants' arousal ratings of positive valence-low arousal, positive valence-medium arousal, and neutral valence-medium arousal categories were affected by temperature conditions. The warm condition ratings were higher than the cool condition in the mentioned conditions. The trends are shown in Figure~\ref{fig:valence_arousal}.

%\begin{figure}
%\centering
%\includegraphics[width=.5\textwidth]{Figures/Valence.png}
%\caption{SAM rating responses for valence. The error bars represent the standard error values.}
%\label{fig:valence}
%\end{figure}

The test showed a main effect of temperature on valence rating (F(2,117) = 3.43, p = 0.036). Images were rated lower in the warm condition compared to the neutral condition. There was no statistically significant interaction between image category and temperature on valence ratings (F(12,696) = 0.77, p = 0.638). The summary of ratings can be seen in Figure~\ref{fig:valence_arousal}. Valence and arousal ratings for each image category and the normative ratings for comparison can be found in Table~\ref{tbl:normative responses}.

\subsection{Geneva Emotion Wheel}

Analyzing emotion wheel data is a challenging task mainly because of the absence of a standard approach in psychology for analyzing this type of data. One obstacle in analyzing emotion wheel data is that the data cannot be considered entirely discrete or continuous. Another hurdle is that the circles in proximity to each other share the same valence/control levels. On the other hand, the fact that it is a circle adds to the challenge of defining a point of reference to quantify emotions. This section presents different approaches devised to analyze the emotion wheel data.

The first approach was to analyze the emotions based on the emotion category and intensity as discrete values. This approach divided responses into 65 categories (16~x~4~+~1), corresponding to 16 emotions with four intensity levels and one neutral. Two other levels of categorization were the image category and the temperature conditions (7 x 3). Since some categories had low response frequency, to eliminate outliers, frequency responses that were less than 15 responses (out of 6,720 total) were removed from the data set to decrease low-confidence results. A Chi-Square test was conducted to determine whether there was a significant difference between the emotional responses between image categories and temperature conditions. The results did not indicate a significant difference between temperature conditions for different image categories ($\mbox{$\chi^2 = (244,N=123)=250.98$}, \mbox{p=0.375}$).

Another approach to analyzing emotion wheel data was to define the geometrical metric. Since the emotion wheel benefits from geometry in defining emotions, it seems logical to pursue the same method in evaluating the results. The approach taken here was to analyze the responses in terms of their angle on the wheel, which corresponds to the type of emotion, and the distance from the center of the circle, which corresponds to the intensity of the emotion. Each participant's average response angle was calculated for each image category. Later, the resulting value was continuous, so it can be analyzed using standard parametric statistical methods. The same process was performed on the distance of the responses to calculate the mean distance for each image category for every participant. This way, a mixed-design analysis of variance (ANOVA) was performed on each data set. In this analysis, the temperature condition was the between-subject independent variable. The image category was a within-subject independent variable with angle/distance mean values as the dependent variable. The results did not demonstrate a significant difference between temperature conditions~\mbox{(F(2,122)=0.219, p=0.804)}. A similar method on the distance of the responses was performed, which did not result in a significant difference between temperature conditions~\mbox{(F(2,126)=0.034, p=0.967)}.

\begin{figure*}
\centering
\includegraphics[width=\textwidth]
{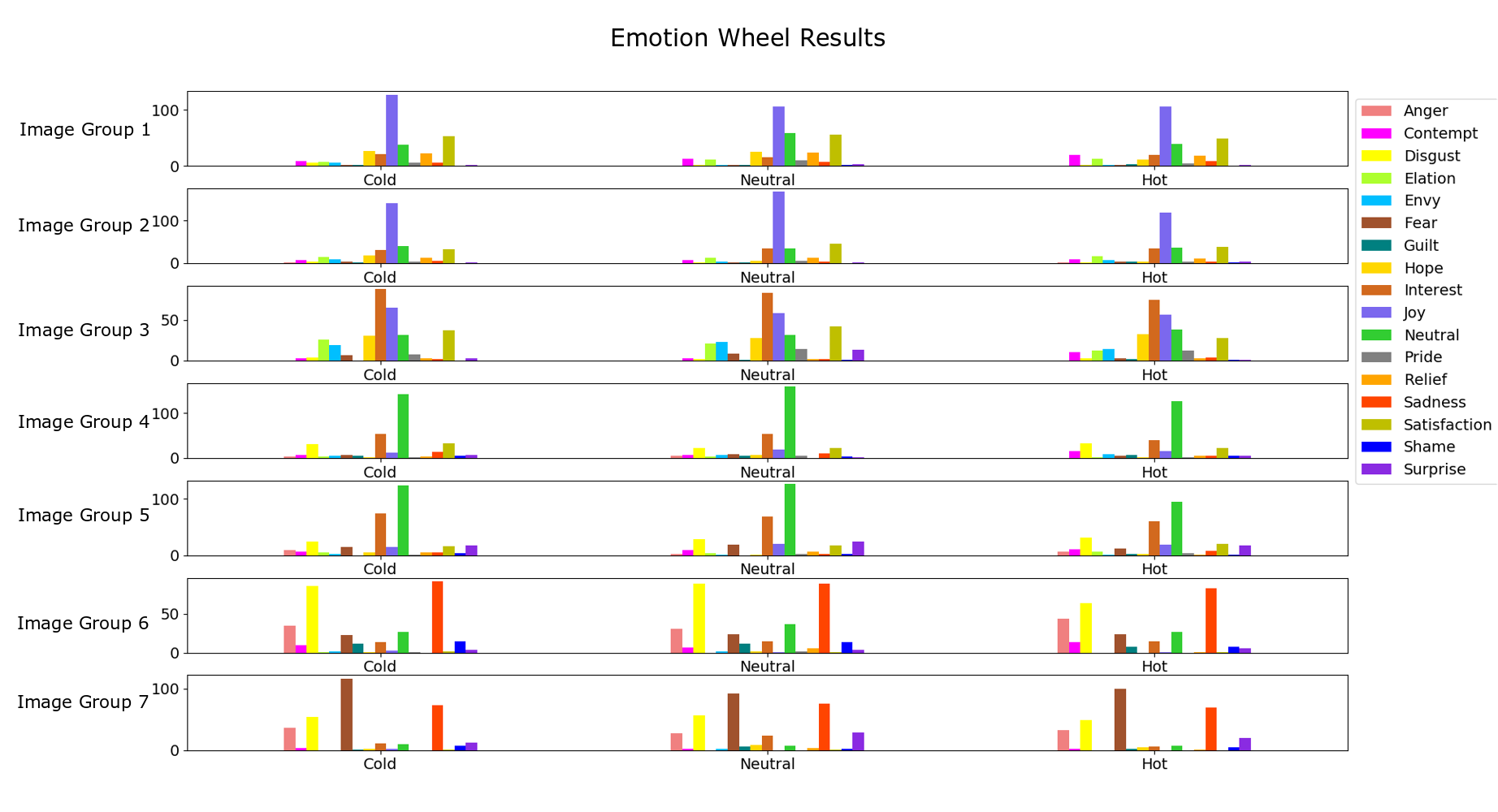}
\caption{Emotion wheel responses for three temperature conditions with a schematic of the emotions and their locations on the emotion wheel. Emotion responses with a frequency response of less than 5 are not included.}
\label{fig:emotionwheelresults}
\end{figure*}

The results of emotion wheel responses are illustrated in Figure~\ref{fig:emotionwheelresults}. On these figures, the frequency of each emotion is proportional to the height of the bars. A chi-square test of independence was performed to examine the relation between temperature conditions and self-reported emotions. The first independent variable was the emotion category, which combined the rated emotion and its image category. The other independent variable was the temperature condition. The dependent variable was the frequency of the responses.

After removing the frequencies that occurred less than 5 times, 54 emotion categories were left with three temperature conditions. Test results did not demonstrate a statistically significant difference between independent variables ($\mbox{$\chi^2 (106, N = 6160) = 110.95$}, \mbox{p = 0.352}$). The results for each temperature condition and emotion are illustrated in Figure~\ref{fig:emotionwheelresults}.

%\begin{figure*}
%\begin{minipage}{6in}
 % \centering
  %\raisebox{-0.5\height}{\includegraphics[width=.8\textwidth]{Figures/emresults.png}}
  %\hspace*{.001in}
  %\raisebox{-0.5\height}{\includegraphics[width=.18\textwidth]{Figures/emotionwheel.png}}
%\end{minipage}
%\end{figure*}
\section{Discussion}

In this study, the contribution of temperature on affective measures was investigated. Naive participants reported their emotions through SAMS after viewing a series of images from the IAPS library while experiencing a cool (30\degree C), neutral (33\degree C), or warm (36\degree C) temperature on the back of their neck. Geneva emotion wheel was used as a complementary method for evaluating affective measures.

In all image categories, the arousal ratings ascended from cool to neutral and neutral to warm conditions. Warm temperature ratings for arousal were significantly higher than the cool temperature in image categories, where the arousal ratings were low or medium (image categories 1, 2, and 5). We hypothesize that the absence of significant difference between high-arousal categories was because the arousal rating for those categories was already high; therefore, the warm temperature needed to increase the affective measure level to the extreme value, which was not plausible with the given temperature conditions. The obtained result agrees with Akazue et al.~\cite{akazue2016effect}, whose findings showed that thermal stimulation increases the arousal rating in images with neutral to low arousal. However, it is worth mentioning that the results obtained in this work are different from the mentioned study regarding the statistical method. Akazue et al. considered the control condition as the normative ratings of the IAPS images, while in this study, the control condition was the neutral temperature condition using a more direct comparison. By comparing the obtained valence-arousal responses and the associated emotion on the emotion circumplex, one can observe that warm temperature moved the affective state from emotions like ``hopeful" to ``feel well". The cool temperature most effectively changed participants' arousal levels, where images elicited neutral arousal emotions. This effect diminishes for unpleasant emotions (categories 1 and 2). Moreover, the difference between valence ratings for neutral and warm conditions is clear evidence of the effect of temperature on affective measures. Warm temperature decreased the pleasantness of the ratings. This difference is highest for image category 2 and can be seen in Figure~\ref{fig:valence_arousal}.

A comparison of results from Figure~\ref{fig:valence_arousal} and Table~\ref{tbl:normative responses} shows that the valence responses were more consistent with normative ratings from the IAPS library for unpleasant and neutral images. The variability of the responses also decreased for valence ratings compared to the arousal ratings. Although the temperature condition affected the valence ratings, overall, this difference was not reflected within each image category. We hypothesize that a more intense thermal stimulus could affect participants' responses on a greater scale. However, there are challenges associated with this approach that require making a trade-off between different aspects of the test design.  The lack of a universal reference point for thermal perception makes temperature a more personal stimulus. Due to the variability of skin temperature, skin type, and metabolism between participants, one can not guarantee a similar thermal perception intensity for all participants. Another challenge is that increasing the temperature intensity could affect the deception method and lead individuals to realize the purpose of the study. Another contributor is the thermal adaptation over the course of the test. Individuals might lose sensitivity to the applied temperature over time, affecting their responses. 

Different exploratory statistical analysis methods are introduced for analyzing the emotion wheel data. The results of the emotion wheel did not demonstrate any effect of temperature on responses. This result does not seem surprising since the images were selected based on their valence and arousal ratings and not by their relation to the same emotional state, e.g., ``anger" or ``sadness". We hypothesize that selecting a more tailored set of images for the emotion wheel can evaluate the role of temperature stimulation on emotions more accurately.

\section{Conclusion}

Results demonstrated that warm temperature increased the arousal ratings of images with low to neutral arousal levels compared to cool temperature. These changes occurred for images with positive valence. The results are consistent with similar studies in the field~\cite{akazue2016effect,halvey2012augmenting,salminen2011emotional}. In these studies, it was shown that warm temperatures led to an increase in arousal ratings. The difference between the previous studies and this study is the location of the thermal actuator and the use of the deception method, consequently minimizing the confirmation bias. The addition of the control condition is another improvement of this study compared to Akazue et al.'s approach~\cite{akazue2016effect}. 

Analyzing the emotion wheel data was challenging since available non-parametric statistical methods are unsuitable for analyzing the collective interaction between all emotions and temperature conditions, given the mixed-design nature of data analysis (presence of within-subject and between-subject variables). Implementing a fully within-subject study design will alleviate these concerns. However, it will increase the experiment length and confirmation bias since participants will need to go through all temperature conditions.

The role of emotions in haptic interactions has been an underinvestigated research topic in the haptics community. As commercialized haptic products become more ubiquitous, studying how each haptic modality can affect users' emotions can unveil the potential of haptic technology. As discussed in this work, the extensive interactions between temperature and emotions show how these two aspects are intertwined and how they can affect our body and mind. Among these, thermal haptic cues seem to have the most direct impact on affective measures to the extent that they could compete with medicinal methods~\cite{hanusch2013whole}. This emphasizes the importance of this line of research and its potential to impact people's lives. Studies similar to the work presented can provide a clearer picture of how to utilize haptic capabilities to increase haptic products' impact.

\section*{Acknowledgments}
This material is based upon work supported by the National Science Foundation under Grant Number \mbox{IIS-1526475}.

\bibliographystyle{IEEEtran}
\bibliography{Temperature_Emotion}

\end{document}